\documentstyle[aps,preprint,draft]{revtex}
\title{Application of Gradient Expansion to Inflationary Universe}
\date{1 Nobember 1994}
\author{Yasusada Nambu and Atsushi Taruya}
\address{Department of Physics, Nagoya University \\ 
         Chikusa, Nagoya 464-01, Japan}
\begin{document}
\maketitle
\begin{abstract}
 Using the long wave perturbation scheme(gradient expansion), the effect 
 of inhomogeneity on the inflationary phase is investigated. We solved the 
  perturbation equation of which source term comes from inhomogeneity of a
   scalar field and a seed metric. The result indicates 
 that sub-horizon scale inhomogeneity strongly affects the onset of inflation.
\end{abstract}
\def\pd#1#2{\frac{\partial#1}{\partial#2}}
\def\fd#1#2{\frac{\delta#1}{\delta#2}}
\def\t0{t_0}
\def\cd#1#2{{#1}_{|#2}}
\def\cu#1#2{{#1}^{|#2}}
\def\k{\kappa}
\def\g{\gamma}
\def\d{\delta}
\def\r3{R(\gamma^{(0)})}

%
\section{Introduction}

Dynamics of inhomogeneity plays an important role in many phenomena 
in general relativity. Generality of inflation is one of such a problem. 
To explain the present our universe, we must have an inflationary phase 
in past. But as we do not know the initial condition of universe, we hope 
that universe can enter inflationary phase from wide range of initial 
condition. 
For homogeneous closed universe, the universe cannot enter inflationary phase
 if the initial curvature is too large\cite{wald,homo}. For inhomogeneous universe,
 the effect of strong(non-linear) initial 
inhomogeneity  is studied for space-time with 
symmetry by numerical method\cite{num}. These result indicate that sub-horizon
 scale inhomogeneity 
strongly prevent the onset of inflationary phase. But imposing symmetry 
to space-time may lose general feature of dynamics of inhomogeneity.
 We want to know about more general situation without 
symmetry using appropriate approximation that include non-linear effect. 
Gradient expansion(GE)\cite{lif,tomi,salo,comer} is one of such a method.
 This approximation scheme 
expands the Einstein equation by the number of the spatial gradient. As 
the background solution, we have an inhomogeneous space-time that does 
not include the effect of a spatial curvature. We can include the 
curvature by calculating next order. This method describes a long-wave 
non-linear perturbation and it is suitable to applying to the inflationary 
universe. Furthermore, we can treat non-linear effect without imposing 
any symmetry for a space-time.

In this paper, we treat Einstein gravity with a minimally coupled scalar 
field by GE. Our purpose is to observe the effect of large scale 
inhomogeneity on inflationary phase. Especially, we are interested in 
how the onset of inflation is affected by the initial inhomogeneity.

This paper is organized as follows.
In Sec.II, we shortly review GE approximation using Hamilton-Jacobi method.
 We derive the second order equation of motion that include the scalar field 
 inhomogeneity. In Sec.III, we obtain the solution for the inflationary phase
 and discuss the effect of inhomogeneity on inflation. Sec.IV is 
devoted to summary.
\section{Equation of motion by H-J method}
We follow the method of Salopek\cite{salo} and derive the second order equation
 of motion of GE by Hamilton-Jacobi(H-J) method. H-J equation is
\begin{eqnarray}
 {\cal H}&=&2\kappa\gamma^{-1/2}\frac{\delta S}{\delta\gamma_{ij}}
   \frac{\delta
     S}{\delta\gamma_{kl}}(\gamma_{jk}\gamma_{il}-\frac{1}{2}
   \gamma_{ij}\gamma_{kl})-\frac{1}{2\kappa}\gamma^{1/2}R+
   \frac{1}{2}\gamma^{-1/2}(\frac{\delta
     S}{\delta\phi})^2+\gamma^{1/2}V(\phi)
   =0, \nonumber \\
 {\cal H}_i&=&-2(\gamma_{ik}\frac{\delta S}{\delta\gamma_{kj}})_{,j}
   +\frac{\delta S}{\delta\gamma_{lk}}\gamma_{lk,i}
   +\frac{\delta S}{\delta\phi}\phi_{,i}=0, 
\end{eqnarray}
where $\k =8\pi G$. These two equations are Hamiltonian constraint and 
momentum constraint, respectively. We expand the generating functional $S$ by
 the number of spatial derivative
$$
 S=S^{(0)}+S^{(2)}+\cdots.
$$
The form of generating functional $S^{(0,2)}$ is assumed to be
\begin{equation}
  S^{(0)}=-\frac{2}{\kappa}\int d^3x\gamma^{1/2}H(\phi), 
  ~S^{(2)}=\int d^3x\gamma^{1/2} (J(\phi)R+K(\phi)\phi_{;i}\phi^{;i}),
\end{equation}
where $;$ represents spatial covariant derivative associated with 
$\gamma_{ij}$.
This form of generating functional satisfies momentum 
constraint strongly provided that parameter fields(integration constant of 
H-J equation) are spatially constant.
 
The zeroth order Hamiltonian becomes
$$
{\cal H}^{(0)}=-\frac{3}{\kappa}\gamma^{1/2}\left[H^2-\frac{2}{3\kappa}
                        \left(\frac{\partial H}{\partial\phi}\right)^2
                        -\frac{\kappa}{3}V(\phi)\right],
$$
and the requirement that it vanishes gives
\begin{equation}
  H^2=\frac{2}{3\kappa}
                        \left(\frac{\partial H}{\partial\phi}\right)^2
                        +\frac{\kappa}{3}V(\phi)
\end{equation}
The second order Hamiltonian becomes
\begin{eqnarray}
{\cal H}^{(2)}=\gamma^{1/2}R\left[HJ-\frac{1}{2\kappa}-\frac{2}{\kappa}
                        \pd{H}{\phi}\pd{J}{\phi}\right]&+&
                \gamma^{1/2}\phi_{;m}\phi^{;m}\left[HK+\frac{1}{2}
                +\frac{2}{\kappa}\pd{H}{\phi}\pd{K}{\phi}-4H\pd{{}^2J}{\phi^2}
                \right]   \nonumber \\ 
                &+&\gamma^{1/2}\phi_{;m}^{~;m}
                \left[\frac{4}{\kappa}\pd{H}{\phi}K-4H\pd{J}{\phi}\right]. 
\end{eqnarray}
The requirement that this vanishes gives the following equations for
$J,K$:
\begin{eqnarray}
 HJ&-&\frac{1}{2\kappa}-\frac{2}{\kappa}H^{'}J^{'}=0, \nonumber \\
 \frac{1}{\kappa}H^{'}K&-&HJ^{'}=0,   \label{hjk}\\
 HK&+&\frac{1}{2}+\frac{2}{\kappa}H^{'}K^{'}-4HJ^{''}=0, \nonumber
\end{eqnarray}
where ${}^{'}=\pd{}{\phi}$. We have two equations for $K$, but it can 
be shown that the second equation automatically satisfies the third one.
So they are consistent.

Now the evolution equation that is accurate to the second order spatial 
gradient becomes the following form. From now on,
 we use the synchronous coordinate system $N=1, N_i=0$:
\begin{eqnarray}
\dot\phi&&=\gamma^{-1/2}\fd{S}{\phi} \nonumber \\
        &&=-\frac{2}{\kappa}\pd{H}{\phi}+\pd{J}{\phi}R(\gamma)-
         \left[\pd{K}{\phi}\phi_{;m}\phi^{;m}+2K\phi_{;m}^{~;m}\right], 
         \label{ev1} \\         
\dot\gamma_{ij}&&=
   4\kappa\gamma^{-1/2}(\gamma_{ik}\gamma_{jl}
   -\frac{1}{2}\gamma_{ij}\gamma_{kl})\fd{S}{\gamma_{kl}} \nonumber \\
   &&=4\kappa (\frac{1}{2\kappa}H\gamma_{ij}
     +K(\frac{1}{4}\gamma_{ij}\phi_{;m}\phi^{;m}-\phi_{;i}\phi_{;j})
      \nonumber \\
      &&~~~~~~~+J(\frac{1}{4}\gamma_{ij}R(\gamma)-R_{ij}(\gamma))
        +J_{;ij})  \label{ev2}
\end{eqnarray}
The first order solution is obtained by integrating the following equation:
\begin{eqnarray}
  \dot\phi&=&
             -\frac{2}{\k}H'(\phi),  \nonumber \\
  \dot\g_{ij}&=&2H\g_{ij},
\end{eqnarray}
and the solution becomes
$$
\gamma^{(0)}_{ij}=A(\phi)h_{ij}(x), ~~\phi^{(0)}=\phi_0(t-\t0(x)),
$$
where $h_{ij}(x)$ is an arbitrary spatial function(seed metric) and 
$A^{1/2}=\int dt H$. $\t0$ is a integration constant that depends only on 
spatial coordinate, and a scale factor $A$ depends on spatial coordinate through 
$\phi$.  We can rewrite above 
 equations (\ref{ev1}), (\ref{ev2}) using $\t0$ and covariant derivative 
 associated with a seed metric 
 $h_{ij}$. The spatial curvature becomes
\begin{eqnarray}
 R_{ij}(\gamma^{(0)})=
   R_{ij}(h)&-&h_{ij}\left[(\dot H+H^2)\cd{\t0}{m}\cu{\t0}{m}-H\Delta\t0
                    \right] \nonumber \\
            &+&(-\dot H+H^2)\cd{\t0}{i}\cd{\t0}{j}+H\cd{\t0}{ij} , \\
 R(\gamma^{(0)})= \frac{1}{A}
   (R(h)&+&(-4\dot H-2H^2)\cd{\t0}{m}\cu{\t0}{m}+4H\Delta\t0), \nonumber
\end{eqnarray}
where ${\t0}_{|ij}$ means covariant derivative associate with $h_{ij}$.
 The derivative of $\phi,J$ is
\begin{eqnarray}
\phi_{;ij}&=&
   \frac{2}{\k}H'\cd{\t0}{ij}-\frac{2}{\k}HH'h_{ij}{\t0}_{|m}{\t0}^{|m}
   +\left(\frac{2}{\k}\right)^2(H''H'+\k HH'){\t0}_{|i}{\t0}_{|j}, \\
4\k J_{;ij}&=&
 8\left[((H')^2+\frac{3}{2}\k H^2)J-\frac{3}{4}H\right]\cd{\t0}{i}\cd{\t0}{j}
 +8(\frac{\k}{2}HJ-\frac{1}{4})\cd{\t0}{ij} \nonumber \\
 & &~~~~~~~~~~-8H(\frac{\k}{2}HJ-\frac{1}{4})h_{ij}\cd{\t0}{m}\cu{\t0}{m}.
\end{eqnarray}
To derive these expression, we have used eq.(\ref{hjk}). 
Using these relations, the form of evolution equation becomes
\begin{eqnarray}
\dot\phi&=&-\frac{2}{\k}H'+\frac{J'}{A}R(h)-\frac{1}{\k}\frac{H'}{A}
             \cd{\t0}{m}\cu{\t0}{m}, 
                  \label{evo1}
                   \\
\dot\g_{ij}&=&
    2H\g_{ij}+4\k J(\frac{1}{4}R(h)h_{ij}-R_{ij}(h))    \label{evo2} \\
    &&~~~~+2H(-\cd{\t0}{i}\cd{\t0}{j}+\frac{1}{2}h_{ij}\cd{\t0}{m}\cu{\t0}{m})-2\cd{\t0}{ij}.
    \nonumber
\end{eqnarray}        

We consider the  perturbation due to the second order spatial gradient. By 
linearizing eq.(\ref{evo1}) and (\ref{evo2}), we get
\begin{eqnarray}
 \delta\dot\phi&=&-\frac{2}{\k}H''\delta\phi+E(\phi_0,h), \label{per1} \\
 \delta\dot h_{ij}&=&2H'h_{ij}\delta\phi+F_{ij}(\phi_0,h),
                 \label{per2}
\end{eqnarray}
where $\d h_{ij}=\d\g_{ij}/A$ and $E,F_{ij}$ are source terms from the 
 second order gradient:
\begin{eqnarray}
 E&=& \frac{J'}{A}R(h)
             -\frac{1}{\k}\frac{H'}{A}\cd{\t0}{m}\cu{\t0}{m},      \nonumber \\
 F_{ij}&=&4\k\frac{J}{A}(\frac{1}{4}R(h)h_{ij}-R_{ij}(h))  
    +2\frac{H}{A}(-\cd{\t0}{i}\cd{\t0}{j}+\frac{1}{2}h_{ij}\cd{\t0}{m}\cu{\t0}{m})
    -\frac{2}{A}\cd{\t0}{ij}.
    \nonumber
\end{eqnarray}
These terms represent non-linear inhomogeneity due to the seed metric 
$h_{ij}$ and the zeroth order matter field $\phi(t-\t0(x))$.
 The solution of (\ref{per1}) becomes
\begin{equation}
  \delta\phi=\left(\int_{ti}^{t} dt\frac{E}{\dot\phi_0}\right)\dot\phi_0, \label{sol1}
\end{equation}
where we have chosen the integration constant such as $\delta\phi(t_i)=0$.
Using this solution, the metric perturbation becomes
\begin{eqnarray}
 \d h_{ij}&=&\int_{ti}^t\left(2H'\delta\phi h_{ij}+F_{ij}\right) \nonumber \\
    &=&2H\int_{ti}^t dt\frac{E}{\dot\phi_0}h_{ij}+\int_{ti}^t dt
    \left(-2H\frac{E}{\dot\phi_0}h_{ij}
         +F_{ij}\right). \label{sol2}
\end{eqnarray}

\section{Solution in the Inflationary phase}
We derive the solution in the inflationary phase $\ddot a>0$. 
We first evaluate the function $J$:
\def\hi{H^{-1}}
\begin{eqnarray}
 J&=&\frac{1}{2\k a}\left(\int dt a+ const\right) \nonumber \\
  &=&\frac{1}{2\k}\{\hi-\hi(\hi)^{\cdot}+\hi(\hi(\hi)^{\cdot})^{\cdot}-\cdots 
  \nonumber \\
   & &~~~~~~
    -\frac{1}{a}\left(a_i\hi_i-a_i\hi_i(\hi_i)^{\cdot}+\cdots\right)+\frac{const}{a}\},
\end{eqnarray}
and we choose the integral constant so that the term proportional to 
$1/a$ vanishes. 
\begin{equation}
 J=\frac{1}{2\k}\hi\left\{1-(\hi)^{\cdot}
         +(\hi(\hi)^{\cdot})^{\cdot}-\cdots\right\}. \label{j} 
\end{equation}
During inflation, we can neglect all but the first term in this 
expression because the condition $|\dot H|<H^2$ is satisfied. 
 To the leading order of this expansion, we can evaluate the 
 following integrals:
\begin{eqnarray}
  \int_{ti}^{t} dt\frac{J'}{\dot\phi_0A}&\approx&
       -\frac{1}{8}\left[a^{-2}H^{-3}\right]_{ti}^t,  \nonumber \\
  \int_{ti}^t dt\frac{H'}{\dot\phi_0A}&\approx&
       \frac{\k}{4}\left[a^{-2}H^{-1}\right]_{ti}^t, \nonumber
\end{eqnarray}
and the perturbation of the scalar field becomes
\begin{equation}
  \frac{\d\phi}{\dot\phi_0}=-\frac{1}{8}R(h)\left[a^{-2}H^{-3}\right]_{ti}^t
          -\frac{1}{4}\cd{\t0}{m}\cu{\t0}{m}\left[a^{-2}H^{-1}\right]_{ti}^t.
\end{equation}
Therefore during inflation, $\d\phi /\dot\phi_0$ approaches a constant
 value  and $\d\dot\phi$
  goes to  zero in a Hubble time scale:
\begin{equation}
 \frac{\d\phi}{\dot\phi_0}\approx\left(\frac{1}{8}R(h)(a^{-2}H^{-3})_{ti}
                 +\frac{1}{4}\cd{\t0}{m}\cu{\t0}{m}(a^{-2}H^{-1})_{ti}\right)
                 \equiv\left(\frac{\d\phi}{\dot\phi_0}\right)_0, \label{dphi}
              ~~\d\dot\phi\approx 0.
\end{equation}
The metric perturbation during inflation is expressed as
\begin{equation}
  \d h_{ij}=2H\left(\frac{\d\phi}{\dot\phi_0}\right)_0h_{ij}
     -(a^{-2}H^{-2})_{ti}R_{ij}(h)
       -(a^{-2})_{ti}{\t0}_{|i}{\t0}_{|j}-(a^{-2}H^{-1})_{ti}
       {\t0}_{|ij}.        \label{dh}
\end{equation}
The metric that is not proportional to $h_{ij}$ approaches to constant 
values in a Hubble time scale. GE is correct if $|\phi_0|>|\d\phi|$ and 
$|h_{ij}|>|\d h_{ij}|$ are satisfied. These conditions are rewritten to be
\begin{eqnarray}
  &&\left|\frac{R(h)}{2a^2H^2}+\frac{{\t0}_{|m}{\t0}^{|m}}{a^2}\right|<2, 
  \nonumber \\
  &&|\r3|<|R_*|\equiv12H^2.   \label{con}
\end{eqnarray}
They are well satisfied during inflation provided that they are satisfied at onset of 
inflation because the quantities in the left hand side of the condition decrease in time 
during inflation. Therefore the expression of perturbation (\ref{dphi}), (\ref{dh}) are
 correct in whole time of inflation.

\subsection{Effect of the spatial curvature on inflation}
We investigate the effect of the curvature inhomogeneity. For this purpose, 
let us consider a local Hubble radius and a local scale factor that are 
defined by
\begin{eqnarray}
 \tilde H&=&\frac{1}{6}\gamma^{ij}\dot\gamma_{ij},  \nonumber \\
 \tilde a&=&\exp(\int dt\tilde H). 
\end{eqnarray}
Using eq.(\ref{evo1}), these quantity become
\begin{eqnarray}
  \tilde H&=&H\left(1-\frac{1}{12H^2}R(\gamma^{(0)})\right), \nonumber,\\
  \tilde a&=&a\left(1+\frac{1}{24}\left[\frac{R(\gamma^{(0)})}{H^2}\right]_{ti}^t
             \right)
\end{eqnarray}
The co-moving Hubble radius becomes
\begin{equation}
  (\tilde a\tilde H)^{-1}=(aH)^{-1}
              \left[1-\frac{1}{24}\left(\frac{R(\gamma^{(0)})}{H^2}
              +(\frac{R(\gamma^{(0)})}{H^2})_{ti}\right)\right]^{-1}.
\end{equation}
This quantity decreases during inflation and grows in non-inflationary phase.
 We choose $t_i, t_f$ are onset and end time of inflation, respectively. 
 At $t=t_i$, $(aH)^{-1}$ has maximum and at $t=t_f$ minimum. The maximum 
 value at $t=t_i$ for $\r3<0$ becomes smaller than  one for $\r3=0$ and 
 larger for $\r3>0$. This implies that the negative $\r3$ enhances and the 
 positive $\r3$ suppresses inflatinary phase. Especially if $\r3$ becomes 
 as large as $R_*=12H^2_{ti}$, the co-moving Hubble radius goes to 
 infinity at $t=t_i$ and inflation may not occur. 
 Of course we cannot apply GE to such a regime because the condition 
 (\ref{con}) is violated. But we extrapolate the behavior for small $|R|$ 
 and predict the behavior for large $|R|$.
 As the curvature 
 radius is defined by $L_R=\sqrt{6/R}$, this critical value corresponds 
 to the scale $L_R=\sqrt{2}H^{-1}_i$. If the scale of curvature 
 inhomogeneity is larger than this scale, inflation may occur. But the 
 scale is smaller than this value, we may not have inflation. This result is 
 consistent with the result of the calculation of numerical relativity\cite{num}.
 
 To see the effect of curvature from a different view point, we consider 
 local e-folding time of inflation:
\begin{equation}
  \tilde N(t_i\rightarrow t_f)=\int_{ti}^{tf}dt{\tilde H}
          \approx N+\frac{1}{24}\left[\frac{\r3}{H^2}\right]_{ti}^{tf}
          \approx N-\frac{1}{24}(\frac{\r3}{H^2})_i,
\end{equation}
where $N$ is the e-folding for $\r3=0$. Positive $R$ makes e-folding 
shorter and negative $R$ makes it longer than $R=0$ case. This also 
indicates that positive spatial curvature suppresses inflationary phase.

We check the strong energy condition:
\begin{equation}
 \rho+\sum p_i=-\frac{6}{\k}H^2\left(
       1+\frac{\dot H}{H^2}(1+\frac{R(h)}{12(a^2H^2)}
         +\frac{1}{6a^2}{\t0}_{|m}{\t0}^{|m})\right).
\end{equation}
At the onset of inflation, $(\dot H/H^2)_{ti}=-1$, so $(\rho+\sum p_i)_{ti}=
H^2/\k(R(h)/(2a^2H^2)_{ti}+{\t0}_{|m}{\t0}^{|m}/(a^2)_{ti})$. Positive 
 $R(h)$ or ${\t0}_{|m}{\t0}^{|m}$ term make $(\rho+\sum 
p_i)>0$ and the strong energy condition is satisfied. This means that 
inflation does not occur. Negative $R(h)$ can make $(\rho+\sum 
p_i)_{ti}<0$ if its value is larger than the 
${\t0}_{|m}{\t0}^{|m}$ term, and the strong energy condition is violated.


\subsection{Numerical calculation}
To check the result of the previous subsection, we solve eq.(13),(14) 
numerically. We prepare the initial condition at pre-inflationary phase. 
For simplicity, we consider only $\t0=0$ case. The model is 
$V(\phi)=1/2m^2\phi^2$. Parameter and initial value is $m^2=0.5, \phi(0)=7,
\dot\phi(0)=-5,J(0)=0$ in $\k=1$ unit. The scalar curvature 
$R(h)=-30,0,30$. This value is chosen to satisfy the condition 
(\ref{con}). Fig.1 is a phase space diagram 
$(\phi,\dot\phi)$. The solid line is $R(h)=0$, dotted line is $R(h)<0$ 
and dashed line is $R(h)>0$. The zeroth order solution($R(h)=0$) has 
inflationary phase from $\phi\approx 6$ to $\phi\approx 1$. For positive 
$R(h)$, the trajectory enters inflationary phase with smaller $\phi$ 
value(smaller $H$). For negative $R(h)$, the trajectory enters inflation with
 larger $\phi$ value(larger $H$). After entering inflation, all trajectories approaches 
attractor solution. Fig.2 is time evolution of the perturbation of scalar field 
$\d\phi$. In Hubble time scale, its value approaches a constant value.
Fig.3 shows the co-moving Hubble radius. $R(h)=0$ curve has a maximum at 
$t\approx 0.06$ and after this time, the system enters inflation. The 
positive $R(h)$ makes the co-moving Hubble radius larger and the negative 
$R(h)$ makes smaller. These behavior is consistent with the analysis of 
the previous subsection.
\section{Summary}

As we have shown, GE becomes good approximation in the inflationary phase 
because the perturbation goes to a constant value. If the initial 
inhomogeneity is not so large as to break the approximation, we can use 
GE as a model to construct an inhomogeneous inflationary space-time. As 
is well known, a global structure of inflation is very complicated 
because of continuous creation of new inflationary domain via quantum fluctuation. 
 We can grasp 
these feature by using a stochastic approach\cite{sto1,sto2,sto3}, but this method does not 
incorporate the information of geometry completely. GE can describe the 
space-time of inflation whose local Hubble radius is different from place 
to place. The zeroth order solution has the form $\phi=\phi_0(t-\t0(x))$ and 
choosing the time delay function $\t0(x)$ appropriately, we can construct 
such a space-time as the solution of GE. We will discuss on 
this topics in a separated publication\cite{next}.
\vskip 1.5cm
{\it Note:}
After this work has completed, we noticed that the paper\cite{deru}. They treat
 this subject in the same context as ours. But they perform only numerical analysis 
 and does not derive analytic expression of perturbation.


\begin{figure}
\caption{A phase space diagram $(\phi,\dot\phi)$. for $R(h)=0$(solid line), 
$R(h)>0$(dotted line) and $R(h)<0$(dashed line). $R(h)=0$ solution enters inflation 
at $\phi\approx 6$. For positive $R(h)$, solution enters inflation with smaller $\phi$
 value and for negative $R(h)$, solution enters inflation with larger $\phi$ value.}
\label{fig1}
\end{figure}
\begin{figure}
\caption{Time evolution of perturbation $\d\phi$. Perturbation approaches constant value
 in Hubble time scale.}
\label{fig2}
\end{figure}
\begin{figure}
\caption{Time evolution of co-moving Hubble radius. System enters inflationary phase at
 $t\approx 0.06$ for $R(h)=0$ case. For positive $R(h)$, co-moving Hubble radius at 
 the onset of inflation becomes larger. For negative $R(h)$, co-moving Hubble radius at
  the onset of inflation becomes smaller.}
\label{figure3}
\end{figure}

\end{document}